\documentclass[aps,prl,preprintnumbers,nofootinbib,twocolumn,showpacs]{revtex4}
\usepackage{bm}
\usepackage{latexsym}
\usepackage{dcolumn}
\usepackage{amsmath,amsfonts,amssymb}
\usepackage{graphicx,epsfig}
\usepackage{psfrag}
\usepackage{amsthm}

\def\be {\begin{equation}}
\def\ee {\end{equation}}
\def\bea {\begin{eqnarray}}
\def\eea {\end{eqnarray}}
\def\bc {\begin{center}}
\def\ec {\end{center}}
\def\bfg {\begin{figure}}
\def\efg {\end{figure}}
\def\bi {\begin{itemize}}
\def\ei {\end{itemize}}
\def\nn {\nonumber}

\def\ri {\right}
\def\pa {\partial}

\def\no {\noindent}

\def\f {\phi}

\def\th {\theta}

\def\vph {\varphi}

\def\beq{\begin{equation}}
\def\eeq{\end{equation}}
\def\br{\begin{eqnarray}}
\def\er{\end{eqnarray}}
\newcommand{\eel}[1] {\label{#1}\end{equation}}
\def\renyi{R\'enyi~}
\newcommand{\bdm}{\begin{displaymath}}
\newcommand{\edm}{\end{displaymath}}

\begin{document}
\title{Entanglement entropy in all dimensions}

\author{Samuel L.\ Braunstein$^1$} \email[email: ]{schmuel@cs.york.ac.uk}
\author{Saurya Das$^2$} \email[email: ]{saurya.das@uleth.ca}
\author{S.\ Shankaranarayanan$^3$} \email[email: ]{shanki@iisertvm.ac.in}

\affiliation{$^1$~Computer Science, University of York, York Y010 5GH, UK}
\affiliation{$^2$~Theoretical Physics Group, University of Lethbridge,
   4401 University Drive, Lethbridge, Alberta, Canada T1K 3M4}
\affiliation{$^3$~School of Physics, Indian Institute of Science
   Education and Research, CET Campus, Thiruvananthapuram 695016,
   India}

\begin{abstract}
It has long been conjectured that the entropy of quantum fields
across boundaries scales as the boundary area. This conjecture has not
been easy to test in spacetime dimensions greater than four because
of divergences in the von Neumann entropy. Here we show that the
R\'enyi entropy provides a convergent alternative, yielding a
quantitative measure of entanglement between quantum field theoretic
degrees of freedom inside and outside hypersurfaces. For the first time,
we show that the entanglement entropy in higher dimensions is
proportional to the higher dimensional area. We also show that the R\'enyi
entropy diverges at specific values of the R\'enyi parameter $q$ in
each dimension, but this divergence can be tamed by introducing a mass
to the quantum field.
\end{abstract}

\maketitle


Entanglement, a term first coined by Schr\"odinger, is an intriguing
and quintessentially quantum mechanical property, which correlates
microscopic systems in a precise way, even if they are separated by
large distances. On the one hand it gives rise to apparent
contradictions (such as the EPR paradox) and on the other, hides
enormous untapped resources for computation and communication (e.g.,
via teleportation). A mathematically precise way of measuring entanglement
has remained elusive however, except in the simplest cases where the
combined system is in a pure state, i.e., for which all quantum numbers
are known. Usually, the entanglement entropy is computed as the von Neumann
entropy associated with $\rho$: $S_{_{\rm vN}} = -{\rm tr}\left( \rho
\ln \rho \ri)$.  For recent reviews, see
\cite{2009-Horodecki.etal-RMP,2010-Eisert.etal-RMP}.

Entanglement as a source of entropy has also been examined in an entirely
different context, that of the microscopic origin of black hole
entropy. The basic idea is that the long-range entanglement of the
quantum fields across a black-hole horizon can leave its mark on
the reduced density matrix of the external degrees of freedom (DOF)
which, in turn, accounts for the black-hole entropy.  Starting with
\cite{1986-Bombelli.etal-PRD,1993-Srednicki-PRL}, it was
demonstrated that to leading order, and for a scalar field in its
global ground state, the entanglement entropy between its DOF
inside and outside a black hole horizon is proportional to the area
of the horizon.  Recently, it was demonstrated that the DOF at or near
the horizon contribute to most of this entropy
\cite{2007-Das.Shankaranarayanan-CQG} and that excited states of the
field lead to sub-leading order power law corrections
\cite{2006-Das.Shankaranarayanan-PRD,2008-Das.etal-PRD}.

However, entanglement as a source of black-hole entropy has a couple
of drawbacks: (i) The proportionality constant depends on the ultra-violet
cut-off and the number of fields present 
(also these are in general independent of each other, although it was recently 
heuristically argued that the requirement of the stability of the cosmos 
relates the two and naturally relates the cut-off to the Planck mass
\cite{2009-Brout-IJMPD}), and
(ii) it is not evident that the area proportionality holds for 
higher spacetime dimensions, say
$D+2 > 4$ (throughout $D+1$ denotes the number of {\it spatial\/}
dimensions)?

Although there have been attempts in the literature
\cite{2004-Calabrese.Cardy-JSTAT,2006-Fradkin.Moore-PRL,2009-Casini.Huerta-JPA,2005-Plenio.etal-PRL},
however, it is yet to be shown from first principles. For instance, in
Refs.
\cite{2004-Calabrese.Cardy-JSTAT,2006-Fradkin.Moore-PRL,2009-Casini.Huerta-JPA},
the authors attempt to obtain universal expressions for higher
dimensions from two-dimensional (conformal field theory) entropy
$c$-functions. It is also interesting to point that in all these
references
\cite{2004-Calabrese.Cardy-JSTAT,2006-Fradkin.Moore-PRL,2009-Casini.Huerta-JPA}, it is implicitly assumed that Srednicki's analysis is extendable to
all dimensions (see page~23 of Ref.~%
\cite{2004-Calabrese.Cardy-JSTAT}). Following Srednicki
\cite{1993-Srednicki-PRL}, if one regularizes the entropy function by
introducing a radial lattice, the sum of partial waves does not
converge and the entropy turns out to be infinite in higher dimensions
\cite{2006-Riera.Latorre-PRA}. In Ref. \cite{2005-Plenio.etal-PRL}, the convergence of the eigenvalues (and,
hence, the entropy) assumes that the parameter $c$ is less than $1/(2
d)$ where $d$ is the number of space dimensions. For higher
dimensional black-holes, the brick-wall entropy contains extra
divergent terms other than from the ultra-violet modes
\cite{2008-Sarkar.etal-PRD}.  By using a different measure of entropy,
we show that a {\it non-divergent} entropy-area relation can be obtained
for all dimensions, and furthermore, that the divergences are similar
in nature to the infrared divergences in QED, which can be tamed by
introducing a mass to the field.

While the von Neumann entropy is the most common measure of entanglement,
it is neither the most general, nor unique. There are other measures,
such as the R\'enyi and Tsallis entropies, which under certain limits
reduce to the von Neumann entropy. In
this work we study entanglement via the R\'enyi entropy defined as
\bea
S^{(q)} \equiv \frac{1}{1-q} \ln \biggl( \sum_{i=1}^{n} p_i^q\biggr).
\label{eq:renyi}
\eea
In the limit that $q \rightarrow 1$, the R\'enyi
entropy reduces to its von Neumann counterpart. Also, like von Neumann
entropy (and unlike Tsallis entropy) R\'enyi entropy is additive and
has maximum value $\ln(n)$ for $p_i = 1/n$. In the framework of
statistical mechanics, R\'enyi entropy may be physically interpreted as
the $q$-derivative of the Free energy with-respect-to
temperature \cite{2011-Baez-Arx}.

Consider a free massless real scalar field propagating in
a $(D + 2)$-dimensional flat spacetime with action
{\small
\bea
\frac{1}{2}  \int dt \, dr \, r^{D}
d\Omega_{_{D}} (\eta^{a b} \pa_{a}\Phi~\pa_{b}\Phi
+g^{\theta_n\theta_n} \,
\pa_{_{\theta_n}}\!\Phi \, \pa_{_{\theta_n}}\!\Phi),
\label{act-gensph}
\eea
}
where the Latin indices ($a$ and $b$) take the values $t$ and $r$,
$\Omega_{D}$ is the $D$-dimensional solid angle, $\eta^{a b}$
is the Minkowski metric, and $g^{\theta_n\theta_n}$ are the
metric coefficients for the angular
coordinates. Decomposing the scalar field in terms of real
hyper-spherical harmonics one has
$$
\Phi (x^{\mu}) = \sum_{\ell m_i} r^{D/2} \,
\vph_{_{\ell m_i}}(t,r)\, Z_{\ell m_i} (\th, \f_i),
~~ 1\leq i\leq D-1. 
$$
The Hamiltonian is given by
$H=\sum_{\ell, m_i} H_{\ell, m_i}$ where
\begin{eqnarray}
H_{\ell, m_i} &=& \frac{1}{2} \int_{0}^{\infty} \!\! dr
\, \biggl\{\pi_{_{\ell m_i}}^2
+ \frac{ \ell (\ell + D - 1) }{r^2} \, \vph_{_{\ell m_i}}^2 \nonumber \\
&&\phantom{\frac{1}{2} \int_{0}^{\infty} \!\! dr
\, \biggl\{ } +  r^{D} \Bigl[\pa_{r}
\Bigl( \frac{1}{r^{D/2}}\,\vph_{_{\ell m_i}}\Bigr)\Bigr]^2 \biggr\},
\label{schwD2-Ham}
\end{eqnarray}
%
%
and where $\pi_{\ell m_i} = \pa_{t} \vph_{\ell m_i}$ is the canonical
momentum. Note that the Hamiltonian of a massless scalar field
propagating in a general (i.e., not necessarily flat) spherically
symmetric spacetime at fixed Lemaitre time coordinate reduces to the
above Hamiltonian \cite{2008-Das.etal-PRD}.

For simplicity, we assume that the scalar field is in the ground
(vacuum) state. As mentioned earlier, our interest is in determining
the quantity of information shared by modes across a ``horizon.''
This may be found by eliminating (tracing out) the quantum
degrees of freedom associated with the scalar field inside a spherical
region of radius ${\cal R}$ (the location of our ``horizon''). The
resulting reduced density matrix can then be used to determine the
strength of quantum correlations across the horizon.

It is not possible to obtain a closed form analytic expression
for the density matrix and hence, we need to resort to numerical
methods. In order to do that we take a spatially uniform radial grid,
$\{r_j\}$, with $a = r_{j + 1} - r_j$. To achieve better precision,
we adopt the middle-prescription in discretizing the terms containing
the derivatives, so
{\small
\beq
A(r)\, \partial_{r}[G(r)] \longrightarrow
A(j + {\textstyle \frac{1}{2}})\,  \left[ G(j + 1) - G(j) \right]/a,
\eeq
}
Discretizing the Hamiltonian~(\ref{schwD2-Ham}), and suppressing the
subscripts $l, m_i$, leads to
{\small
\br
\label{disc-SchHam}
H &=& \sum_{j} H_{j} =
\frac{1}{2a} \sum_{i,j}^{N} (\delta_{ij} \pi_{j}^2
+ \vph_{j} \, K_{ij} \, \vph_i ),
\er
}
where the interaction matrix, $K_{ij}$, is given by
{\small
\br
\!\!\!&& \phantom{=\,}\frac{1}{j^D}
\left[(j + {\textstyle \frac{1}{2}} )^D
  + j^{D - 2} \ell (\ell + D - 1)\right] \delta_{j}^i \delta_{i}^1 \nn \\
\!\!\!&&+\,  \frac{1}{j^D}
\left[(j + {\textstyle \frac{1}{2}} )^D \! +
(j - {\textstyle \frac{1}{2}} )^D \! + j^{D - 2} \ell (\ell + D - 1)
\right] \delta_{j}^i \delta_{i}^k \nn \\
\!\!\!&&-\, \frac{\left(j + \frac{1}{2}\right)^D}{j^{D/2}
\, i^{D/2}} \; \delta_{i}^{j + 1}
- \frac{\left(i + \frac{1}{2}\right)^D}{i^{D/2} \, j^{D/2}}
\; \delta_j^{i + 1}\nn \\
\label{eq:Kij}
\!\!\!&&+\, \frac{1}{j^D}
\left[(N - {\textstyle \frac{1}{2}} )^D
+  j^{D - 2} \ell (\ell + D - 1) \right] \delta_{j}^i \delta_{i}^N,
\er
}
$\!\!$and $2 \leq k \leq (N - 1)$. The procedure to obtain the entanglement
entropy in higher dimensions is similar to the one discussed in
Refs.~\cite{1993-Srednicki-PRL,2010-Das.etal-Book}. In this work we
assume that the quantum state corresponding to the Hamiltonian of the
$N$-Harmonic oscillator system (\ref{disc-SchHam}) is the ground state
with wave-function $\Psi_{GS}(x_1,\ldots,x_n;t_1,\ldots,t_{N- n})$. The
reduced density matrix $\rho(\vec t,{\vec t\,}')$ is obtained by tracing
over the first $n$ of the $N$ oscillators
\beq
\int (\Pi_{i =1}^{n} dx_i)\,
\Psi^{*}_{GS}(x_1, \ldots, x_n;\vec t\,)
\Psi_{GS}(x_1, \ldots, x_n;{\vec t\,}') .
\eeq
The \renyi entanglement entropy in $(D + 2)$-dimensional spacetime
is then given by
{\small
\br
\label{eq:renyil}
S^{(q)}(n,N) &=& \sum_{\ell} (2 \ell + D - 1)
\, {\cal W}(\ell) \, S_\ell^{(q)}(n, N),  \\
\label{eq:Weightl} {\cal W}(\ell) &=& \frac{(\ell + D - 2)!}{(D - 1)!
   \;\ell\,!}, 
\er } 
is the angular degeneracy factor
\cite{1997-Kim.etal-PRD,2008-Sarkar.etal-PRD} and $S_\ell^{(q)}(n, N)$
is the R\'enyi entropy for partial waves with total angular momentum
$\ell$.

To highlight the advantage of 
\renyi entropy as the measure
of the entanglement entropy as compared to von Neumann entropy,
we determine the \renyi entropy in the large $\ell \gg N$ limit. In
this limit, Eq.~(\ref{eq:renyil}) becomes
\beq
\label{eq:Renyiasympt}
S^{(q)}(n,N) \simeq
\sum_{\ell} (2 \ell + D - 1)  \frac{{\cal W}(\ell)}{1 - q}
\;\xi_{_{\ell}}^{\,q},
\eeq
where the asymptotic eigenvectors and angular degeneracy factors,
respectively, are
{\small
\begin{eqnarray}
\xi_{_{\ell}} &\simeq&  \frac{1}{2^{2(D + 1)}}
\frac{(2  n + 1)^{2 ( D + 1)} [n (n + 1)]^{3 - D}}{\ell^2 (\ell + D - 1)^2}
\nonumber \\
{\cal W}(\ell) & \simeq & \frac{1}{(D  - 1)!}\,
\Bigl(\frac{\ell}{e}\Bigr)^{D - 2}.
\end{eqnarray}
}
Substituting these expressions into Eq.~(\ref{eq:Renyiasympt}) yields
\beq
\label{eq:Renyiasympt2}
S^{(q)}(n,N) \sim \sum_{\ell} \ell^{D - 1 - 4 q}.
\eeq
In the limit $q \to 1$, Eq.~(\ref{eq:Renyiasympt2}) is identical to
the asymptotic limit of the von Neumann entropy, and diverges
for all $D > 2$ (as also noted in Refs.~%
\cite{1993-Srednicki-PRL,2006-Riera.Latorre-PRA}). By contrast, if
that limit is not taken, 
Eq.~(\ref{eq:Renyiasympt2}) converges for all $q > D/4$. 
For instance, in
4-dimensional spacetime this implies that for all values of
$q > \frac{1}{2}$ entanglement entropy converges, while in
6-dimensions, this entropy converges only for $q > 1$.
Herein lies the main advantage of using R\'enyi entropy. 
The above asymptotic expression also provides an understanding 
as to why von Neumann entropy converges for 4-dimensions and not for
6-dimensions.

We will now show that the \renyi entropy indeed 
provides a good measure for
the entanglement entropy in all dimensions for $q > D/4$, and
also reproduces the 
area proportionality in each case. 
We compute
the \renyi entropy numerically for the discretized Hamiltonian
(\ref{disc-SchHam}). The computations are done using Matlab for
the lattice size $N=300$, $100 \leq n \leq 200$ and the relative
error in the computation of \renyi entropy is $10^{-6}$. The
computations were done for 4-, 5-, 6- and 10-dimensional spacetimes.
However, in this manuscript we only present the results for
5 and 10 dimensions.

\begin{figure}[!htb]
\begin{center}
\includegraphics[scale=0.47]{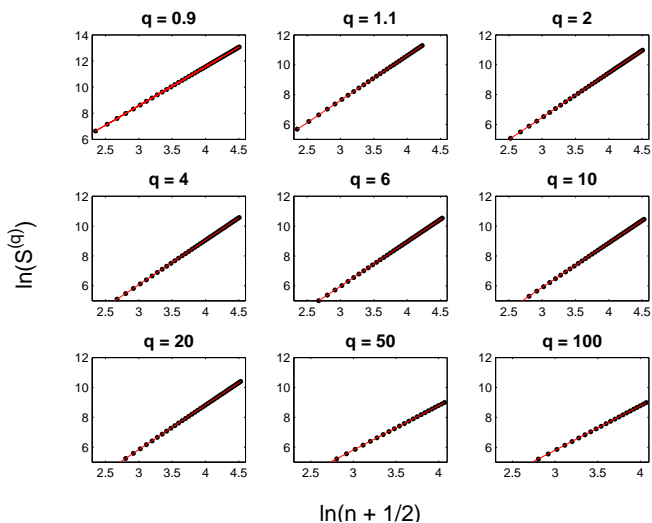}
\caption{Log-log plot of \renyi entropy, $S^{(q)}$, versus the scaled
   radius of the sphere, $s\equiv {\cal R}/a=n + \frac{1}{2}$, for
   $D = 3$, $N = 300$ and $100 \leq n \leq 200$. The
   black dots represent the numerical output and straight lines denote
   lines of best fit. These fits are: $S^{(0.9)} = 0.69 \, s^{2.99}$,
   $S^{(1.1)} = 0.26 \, s^{2.99}$, $S^{(2)} = 0.068 \, s^{2.99}$,
   $S^{(4)} = 0.051 \, s^{2.99}$,
   $S^{(6)} = 0.050 \, s^{2.98}$, $S^{(10)} = 0.047 \, s^{2.98}$,
   $S^{(20)} = 0.045 \, s^{2.98}$, $S^{(50)} = 0.045 \, s^{2.98}$ and
   $S^{(100)} = 0.045 \, s^{2.97}$.}
\label{fig:5D}
\end{center}
\vspace*{-0.05cm}
\end{figure}

\begin{figure}[!htb]
\begin{center}
\includegraphics[scale=0.47]{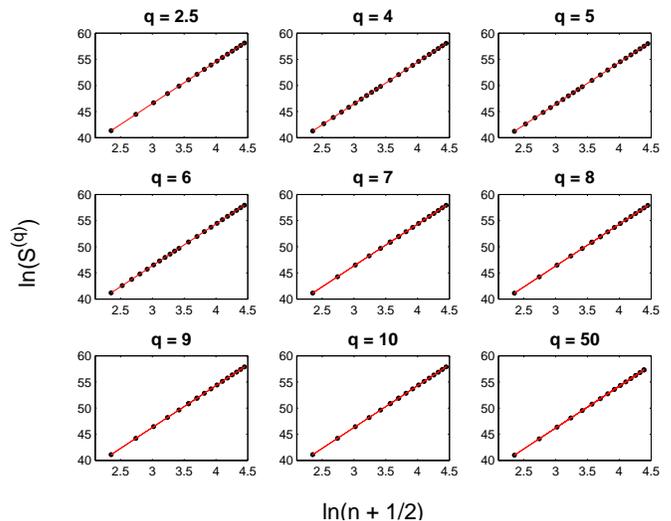}
\caption{Log-log plot of R\'enyi entropy, $S^{(q)}$, versus the scaled
   radius of the sphere, $s\equiv {\cal R}/a=n+\frac{1}{2}$, for $D =
   8$, $N = 300$ and $100 \leq n \leq 200$. The fits are:
   $S^{(2.5)} = 6.08 \times 10^{9} \, s^{8.00}$,
   $S^{(4)} = 5.71 \times 10^{9} \, s^{8.00}$,
   $S^{(5)} = 5.35 \times 10^{9} \, s^{8.00}$,
   $S^{(6)} = 5.14 \times 10^{9} \, s^{8.00}$,
   $S^{(7)} = 5.01 \times 10^{9} \, s^{8.00}$,
   $S^{(8)} = 4.89 \times 10^{9} \, s^{8.00}$,
   $S^{(9)} = 4.81 \times 10^{9} \, s^{8.00}$,
   $S^{(10)} = 4.76 \times 10^{9} \, s^{8.00}$, and
   $S^{(50)} = 4.37 \times 10^{9} \, s^{8.00}$.}
\label{fig:10D}
\end{center}
\vspace*{-0.05cm}
\end{figure}

\begin{figure}[!htb]
\begin{center}
\includegraphics[scale=0.4]{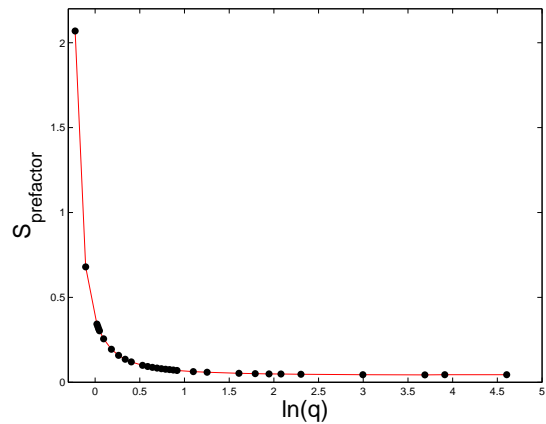}
\caption{Plot of the prefactor, $S_{\text{prefactor}}$, of the \renyi
entropy in 5-dimensional spacetime versus $\ln(q)$. The small circles
represent the numerical output for each run; the red line shows a
spline interpolation. The divergence is apparent as $q\to \frac{3}{4}$.}
\label{fig:qdep}
\end{center}
\vspace*{-0.1cm}
\end{figure}

In Figs.~\ref{fig:5D} and \ref{fig:10D} we have plotted $\ln S^{(q)}$
versus $\ln(R/a)$  for 5- and 10-dimensional spacetimes, respectively.
  From the best fit curves in Fig.~\ref{fig:5D} we see that 
for different values of $q$ the \renyi entropy generically scales
with an approximate power law $S^{(q)} \sim ({\cal R}/a)^3$.
Although as we increase $q$, the prefactor decreases, nonetheless
the power remains close to $3$. Similarly, in the case of
10-dimensional spacetime (Fig.~\ref{fig:10D}), for all $q$,
the power always remains close to $8$. It is important to note that,
although it has long been conjectured in the literature that the
entanglement entropy is proportional to the area of the
$D$-dimensional surface (for $D \geq 3$), 
this is the
first time this has been explicitly shown in higher spatial dimensions.

In Fig.~\ref{fig:qdep} the prefactor $S_{\text{prefactor}}$,
(the ratio of the \renyi entropy to 5-dimensional area) is plotted 
versus $q$, from which we infer the following:
(i) \renyi entropy saturates to a constant value for large $q$.
(ii) Consistent with the asymptotic analysis of
Eq.~(\ref{eq:Renyiasympt2}), the \renyi entropy diverges logarithmically
as $q \to \frac{3}{4}$. Similarly, for $(D+2)$-dimensional spacetime,
it diverges as $q \rightarrow {D}/{4}$.

\begin{figure}[!htb]
\begin{center}
\includegraphics[scale=0.4]{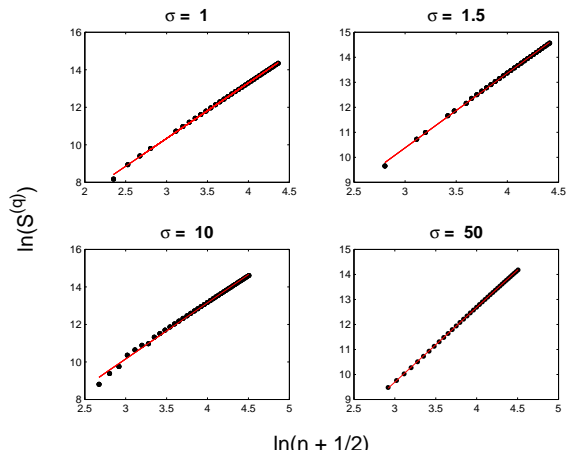}
\caption{Log-log plot of \renyi entropy, $S^{(q)}$, versus the scaled
   radius of the sphere, $s\equiv {\cal R}/a=n+\frac{1}{2}$, in
   5-dimensional spacetime for $q= \frac{3}{4}$ (which is divergent in
   the massless case) and different non-zero scaled masses
   ${\cal \sigma} = m a$. Here $N = 300$ and $100 \leq n \leq 200$.
   The black dots represent the numerical output and straight lines are
   the lines of best fit. These fits are:
   $S^{(3/4)}(\sigma = 1) = 4.25\, s^{2.97}$,
   $S^{(3/4)}(\sigma = 1.5) = 4.16\, s^{2.99}$,
   $S^{(3/4)}(\sigma = 10) = 3.20\, s^{2.99}$ and
   $S^{(3/4)}(\sigma = 50) = 2.07\, s^{2.99}$.  }
\label{fig:mass5D}
\end{center}
\vspace*{-0.25cm}
\end{figure}

To understand this divergence further, we now consider a 
scalar field of mass $m$. The new interaction matrix is
\beq
K_{ij}^{(m)} = K_{ij} + (m \, a)^2 \delta_{ij} \, ,
\eeq
where  $K_{ij}$ is given by Eq.~(\ref{eq:Kij}). 
In Fig.~\ref{fig:mass5D} we plot $\ln S^{(q)}$
versus $\ln(R/a)$ again for 5-dimensional spacetime 
{\it but\/} for $q=\frac34$, and for different values of $m$, and show that 
the divergences disappears!
We also note that as the mass increases, the prefactor
$S_{\text{prefactor}}$ of the \renyi entropy modestly decreases.

To summarize, in this work we have computed \renyi entropy after
tracing out the degrees of freedom inside a $D$-dimensional
sphere of radius ${\cal R}$, and show that the entanglement entropy 
is proportional to the area of the $D$-dimensional surface (for $D \geq 3$).
Although conjectured before, 
this is the first time to our knowledge, that such area proportionality
has been shown explicitly for higher spatial dimensions. 
We have also shown
that the logarithmic divergence of \renyi entropy can be tamed by
considering massive fields.

We conclude with the following points:
\noindent
{\bf 1.} Eq.~(\ref{eq:renyil}) involves summation over partial waves
$\ell$ that extend to infinity. Since the massless Klein-Gordon
field does not have an intrinsic length scale, all angular momentum
modes may contribute. However, for large enough values of
$\ell$, the energy corresponding to the angular momentum modes will
exceed the Planck energy. One way to address 
this difficulty may be to introduce higher-order derivatives to the scalar
field action. However, it is not clear whether the entanglement
entropy-area relation would emerge in this case.

\noindent
{\bf 2.} The entanglement entropy of a massless scalar-field is plagued
by ultra-violet divergences. This is particularly disturbing since it
is a free field theory without interactions, for which one
normally understands how to 
absorb those divergences. Here on the other hand, the
ultra-violet cut-off seems real,
without however any clear physical meaning or knowledge
about how it is implemented
in the real world.

\noindent
{\bf 3.} Over the last decade it has been shown that
higher-dimensional black-holes can have a much richer topological
structure than 4-dimensional black-holes. For instance,
it was shown by Emparan and Reall \cite{2002-Emparan.Reall-PRL} that
horizons with a topology of $S^2 \times S^1$ can exist for
some asymptotically flat spacetimes in 5-dimensions. It would be
interesting to investigate whether the entanglement entropy-area
relation holds for such topologies.

\vskip 0.05truein \no {\it Acknowledgments:} SS thanks S. Theisen for
discussions. SD is supported in part by the Natural Sciences and
Engineering Research Council of Canada and by the Perimeter Institute
for Theoretical Physics. SS is supported by the Department of Science
and Technology, Government of India through Ramanujam fellowship and
Max Planck-India Partner Group in Gravity and Cosmology.

\vskip -0.1truein


\end{document}